\documentstyle[12pt]{article}
\def\sp{\phantom{a}}

\makeatletter
\@addtoreset{equation}{section}

\makeatother

\begin{document}
\sloppy
\sloppy
\sloppy

\begin{flushright}{ UT-881}
\end{flushright}

\vskip 2.5 truecm

\begin{center}
{\Large  A component of superconnection of 11-dimensional curved superspace 
         at second order in anticommuting coordinates}
\end{center}
\vskip 2 truecm
\centerline{\bf Shibusa Yuuichirou}
\vskip .4 truecm
\centerline {\it Department of Physics,University of Tokyo}
\centerline {\it Hongo 7-3-1,Bunkyo-ku,Tokyo 113-0033,Japan}
\vskip 4 truecm

\begin{abstract}

We calculate a component of connection superfields and Lorentz 
superparameter at second order in anticommuting 
coordinates in terms of the component fields of 11-dimensional 
on-shell supergravity by using `Gauge completion'. 
This configuration of superspace holds the $\kappa $-symmetry for 
supermembrane Lagrangian and represents 11-dimensional on-shell supergravity.

\end{abstract}

\newpage

\section{Introduction}
In recent developments of string theory, the deep connection between
supergravity and super Yang-Mills theory has clarified. One remarkable
example is the AdS/CFT correspondence ~\cite{mal}. In this researches,
many supergravity theories which can be obtained by dimensional reduction
from 11-dimensional supergravity have played important roles ~\cite{pop}.

On the other hand, some years ago, T. Banks, W. Fischler, S. H. Shenker
and L. Susskind (BFSS) proposed
that Matrix theory gives a complete description of light-front 
M-theory~\cite{BFSS}. It had been proposed as a theory of D0-branes by 
E. Witten~\cite{w96}. The candidate of its extension on curved
backgrounds is the supermembrane theory. It is described as nonlinear 
sigma model~\cite{ts87} and couples to 11-dimensional superspace
backgrounds that satisfy a number of constraints which are equivalent 
to 11-dimensional on-shell supergravity~\cite{kap}.

Thus, it is important that getting much knowledge of 11-dimensional 
superspace structure. Nevertheless we have little knowledge of it. 
By E. Cremmer, S. Ferrara, L. Brink and P. Howe the on-shell 
conditions and components of superfields up to first order in 
anticommuting coordinates was investigated ~\cite{kap}. By B. de Wit,
K. Peeters and J. Plefka the components of 3-form superfield and part of 
components of vielbein superfields up to second order in 
anticommuting coordinates was investigated ~\cite{wit98}. In the
previous paper, the remaining components of vielbein superfields up to 
second order in anticommuting coordinates was investigated ~\cite{shibu}.

In this paper, we compute part of components of connection superfields 
and the components of Lorentz superparameter at second 
order in anticommuting coordinates in terms of the component fields 
of 11-dimensional on-shell supergravity by using `Gauge completion'.
This configuration of superspace  holds the $\kappa $-symmetry for 
supermembrane Lagrangian and represents 11-dimensional on-shell supergravity.  
 
The paper is organyized as follows. In section 2, we explain `gauge 
completion'. In section 3, we compute parts of 
the superfields. Our conventions are summarized in Appendix.

\section{Gauge Completion}
`Gauge completion' was introduced to identify superspace 
representation as on-shell supergravity~\cite{gc}. In this section we
review this method. 

`Gauge completion' is searching for structures of the superfields and 
superparameters which are compatible with ordinary supergravity. 
That is to say, supertransformations (\ref{TT}) - (\ref{GT}) are
identified as transformations in 11-dimensional spacetime (\ref{st}) and 
the $\theta =0$ components of superfields and superparameters are 
identified as the fields and parameters of ordinary supergravity.

\subsection{Supersymmetry algebra}
Supersymmetry transformations in components formalism are as follows,
\begin{eqnarray}
\label{st}
  \delta _s e_m ^{\sp a} &=      & 2 \bar{\epsilon }\Gamma ^a \psi _m, 
                                  \nonumber \\
  \delta _s \psi _m      &=      & D_m(\hat{\omega })\epsilon + 
                                  T_m ^{\sp rstu}\epsilon \hat{F}_{rstu} 
                                  \equiv \hat{D}_m(\hat{\omega })
                                  \epsilon, \nonumber \\
  \delta _s C_{klm}      &=      & -6 \bar{\epsilon }\Gamma _{[kl}\psi _{m]},\\
  \mbox{with} ,  
  T_m ^{\sp rstu}        &\equiv & \frac{1}{288}(\Gamma _m^{\sp rstu}-8
                                     \delta _m^{[r}\Gamma ^{stu]}),
\end{eqnarray}
where $\hat{F}(= F_{klmn}+12 \bar{\psi}_{[k}\Gamma _{lm}\psi _{n]})$ is 
the supercovariant field strength,
and $\hat{\omega }(= \omega _{m \sp b}^{\sp a}+\frac{1}{2}\bar{\psi _n}
\Gamma ^{\sp a \sp np}_{m \sp b}\psi _p )$ is the supercovariant spin 
connection. And other notation is the same as that in ~\cite{shibu}. 

Its algebra is as follows,
\begin{eqnarray}
 [\delta _{susy}(\epsilon _1),\delta _{susy}(\epsilon _2)] = 
   \delta _g(\xi _3)+\delta _s(\epsilon _3)+
    \delta _l(\lambda _3)+\delta _c(\xi_{3mn}),
\end{eqnarray}
where 
\begin{eqnarray}
\label{alg}
 \xi _3^m     &=& \bar{\epsilon }_2\Gamma ^m
                  \epsilon _1 -(1 \leftrightarrow 2), \nonumber \\
 \epsilon _3  &=& -\bar{\epsilon }_2\Gamma ^n\epsilon _1 \psi _n 
                  -(1 \leftrightarrow 2), \nonumber \\
 \lambda _
 {3\sp b}^{
  \sp a}     &=& -\bar{\epsilon }_2\Gamma ^n\epsilon _1 \hat{\omega }_{n\sp 
                 b}^{\sp a} 
                 +\frac{1}{144}\bar{\epsilon }_2(\Gamma _{\sp b}^{a\sp rstu}
                 \hat{F}_{rstu}+24\Gamma _{rs}\hat{F}_{\sp b}^{a \sp rs})
                 \epsilon_1 -(1 \leftrightarrow 2), \nonumber \\
 \xi _{3mn}  &=& -\bar{\epsilon }_2\Gamma ^k\epsilon _1 C_{kmn}
                 -\bar{\epsilon }_2\Gamma _{mn}\epsilon _1
                 -(1 \leftrightarrow 2). 
\end{eqnarray} 

On the other hand, transformations in superspace formalism are as
follows.
The supertransformation is equal to 
\begin{eqnarray}
\label{TT}
 \delta _T X_{M_p ... M_1} = \Xi ^K \partial _K X_{M_p ... M_1}+p\partial 
                             _{[M_p} \Xi ^K X_{|K|M_{p-1}...M_1]}
\end{eqnarray}
for p-form's components.
The local Lorentz transformations are equal to 
\begin{eqnarray}
\label{LT}
 \delta _L E^A              &=& E^B \Lambda _B^{\sp A},  \nonumber \\
 \delta _L \Omega_B^{\sp A} &=& -\Lambda_B^{\sp C}\Omega _C^{\sp A} +
                                 \Omega _B^{\sp C}\Lambda _C^{\sp A}-
                                d\Lambda _B ^{\sp A} .
\end{eqnarray}
The supergauge transformations are equal to 
\begin{eqnarray}
\label{GT}
  \delta _G B_{LMN} = 3\partial _{[L}\Xi _{MN]} .
\end{eqnarray}
We obtain the full algebra of these transformations as follows
\begin{eqnarray}
 [\delta _T(\Xi _1)+\delta _L(\Lambda _1)+\delta _G
    (\Xi_{1MN}),\delta _T(\Xi _2)+\delta _L(\Lambda 
   _2)+\delta _G(\Xi_{2MN})]  \nonumber \\
    = \delta _T(\Xi _3)+
    \delta _L(\Lambda _3)+\delta _G(\Xi_{3MN}),
\end{eqnarray}
where,
\begin{eqnarray}
\label{ALG}
 \Xi _3^K                   &=& \Xi _2^L\partial _L \Xi _1^K +\delta _1 
                                \Xi _2^K-(1 \leftrightarrow 2), \nonumber \\
 \Lambda _{3A}^{\sp \sp B}  &=& -\Xi _1^K \partial _K\Lambda _{2A}^
                                {\sp \sp B} +\delta _1\Lambda _{2A}^
                                {\sp \sp B} +\Lambda_{1A}^{\sp \sp C}
                                \Lambda _{2C}^{\sp \sp B}
                                 -(1 \leftrightarrow 2), \nonumber \\
 \Xi _{3MN}                 &=& \delta_1\Xi _{2MN}-\Xi _1^K\partial _K
                                \Xi _{2MN} -2\partial _{[M}\Xi _{2N]K}
                                \Xi _1^K -(1 \leftrightarrow 2).          
\end{eqnarray}

\subsection{Gauge completion}
Firstly, we choose the input data as follows
\begin{eqnarray}
 E_m^{\sp a(0)}         &=& e_m^{\sp a}, \nonumber \\
 E_m^{\sp \alpha (0)}   &=& \psi _m^{\sp \alpha }, \nonumber \\
 \Omega _{mb}^
 {\sp \sp \sp a(0)}       &=& -\hat{\omega }_{m \sp b}^{\sp a}, \nonumber \\
 \Xi ^{m(0)}            &=& \xi ^m, \nonumber \\
 \Xi ^{\mu (0)}      &=& \epsilon ^{\mu }, \nonumber \\
 \Xi ^{(0)}_{mn}        &=& \xi _{mn}, \nonumber \\
 B_{mnl}^{(0)}          &=& C_{mnl}.  
\end{eqnarray}
From (\ref{LT}), we obtain 
\begin{eqnarray}
 \Lambda _b ^{\sp a \sp (0)} = \lambda _{\sp b}^a .
\end{eqnarray}
Moreover we introduce the assumption that superparameters do not include the 
derivative of $\epsilon$. 
Then, the higher order components in anticommuting coordinates can be
obtained  by requiring consistency between the algebra of 
superspace supergravity and that of ordinary supergravity.

If we can represent $\Xi_{MN} = 2\partial_{[M}\Phi_{N]}$, we can 
choose the gauge as $\Xi_{MN}=0$ because this superparameters do not 
change the 3-form 
superfields (\ref{GT}) and the algebra (\ref{ALG}). 
Thus we can choose the gauge as follows,
\begin{eqnarray}
 \Xi _{\mu N}^{(0)}    &=& 0 .
\end{eqnarray}

To obtain the higher order components of superparameters  
which depend on $\epsilon$, we must calculate the commutation of 
two supersymmetry transformation.

According to (\ref{alg}),(\ref{ALG}) and (\ref{sam}),
\begin{eqnarray} 
[\delta_{s1},\delta_{s2}]E_m^{\sp a (0)}&=& (\Xi_3 ^K\partial _KE_m^{\sp a}
                                            +\partial _m \Xi_3 ^K E_K^{\sp a}
                                            + E_m^{\sp b}\Lambda_{3b}^{\sp \sp
                                            a})|_{\theta =0} \nonumber \\
                                        &=& (\delta_g(2\bar{\epsilon }_2
                                            \Gamma ^m\epsilon _1)+\delta_s(
                                            -2\bar{\epsilon }_2\Gamma ^n
                                            \epsilon _1 \psi _n)+\delta_c(
                                            -2\bar{\epsilon }_2\Gamma ^k
                                            \epsilon _1 C_{kmn}-2\bar{
                                            \epsilon }_2\Gamma _{mn}
                                            \epsilon _1) \nonumber \\
                                        & & +\delta_l(
                                            -2\bar{\epsilon }_2\Gamma ^n
                                            \epsilon _1 \hat{\omega }_{n
                                            \sp b}^{\sp a}+\frac{1}{72}\bar{
                                            \epsilon }_2(\Gamma _{\sp b}^{a
                                            \sp rstu}\hat{F}_{rstu}+24
                                            \Gamma _{rs}\hat{F}_{\sp b}^{a 
                                            \sp rs})\epsilon_1 ))e_m^{\sp a}.
\end{eqnarray}
Thus one obtains 
\begin{eqnarray}
    \Xi^{k(1)}(susy) = \bar{\theta} \Gamma^k \epsilon.
\end{eqnarray}

In the same way, to obtain the higher order components of superparameters  
which depend on $\lambda$ we must calculate the commutation of 
supersymmetry transformation and Lorentz transformation. 
To obtain the higher order components of superparameters 
which depend on $\xi_{mn}$ we must calculate the commutation of 
supersymmetry transformation and gauge transformation.  
To obtain the higher order components of superparameters which depend 
on $\xi^m$ we must calculate the commutation of supersymmetry transformation
and general coordinate transformation. 
According to superspace algebra,
\begin{eqnarray}
 \delta_{susy} E_m^{
\sp a}|_{\theta =0}    &=& (\Xi ^K(susy)\partial _KE_m^{\sp a}
                           +\partial _m \Xi^K(susy)E_K^{\sp a}+ E_m^{
                           \sp b}\Lambda 
                           _b^{\sp a}(susy))|_{\theta =0} \nonumber \\
                       &=& \epsilon ^{\nu }\partial _{\nu }
                           (E_m^{\sp a (1)})+\partial_m\epsilon ^{\nu }
                           E_{\nu }^{\sp a(0)},  
\end{eqnarray}
while in ordinary supergravity
\begin{eqnarray}
 \delta_{susy} e_m^{\sp a} = 2\bar{\epsilon }\Gamma ^a \psi _m.
\end{eqnarray}
Thus, one obtains 
\begin{eqnarray}
\label{sam}
 E_{\nu }^{\sp a \sp (0)}  &=& 0, \nonumber \\
 E_m^{\sp a (1)}           &=& 2\bar{\theta }\Gamma ^a\psi _m .
\end{eqnarray}
By this procedure, the following results had been 
known ~\cite{kap},~\cite{wit98},~\cite{shibu} .
\begin{eqnarray}
 \Xi ^m        &=& \xi ^m +\bar{\theta }\Gamma ^m\epsilon -\bar{\theta }
                   \Gamma ^n\epsilon \bar{\theta }\Gamma ^m\psi _n
                   +{\cal O}(\theta ^3), \\
 \Xi ^{\mu}    &=& \epsilon ^{\mu }-\frac{1}{4}\lambda _{cd}(\Gamma ^{cd}
                   \theta )^{\mu }-\bar{\theta }\Gamma ^n\epsilon 
                   \psi _n ^{\sp \mu }+\bar{\theta}\Gamma^n\epsilon\bar{\theta}
                   \Gamma^k\psi_n\psi_k^{\sp \mu}+\frac{1}{4}\bar{\theta}
                   \Gamma^n\epsilon\hat{\omega}_{nab}(\Gamma^{ab}
                   \theta)^{\mu}  \nonumber \\
               & & -\frac{1}{3}\bar{\theta}\Gamma^k\epsilon(T_k^{\sp abcd}
                   \theta)^{\mu}\hat{F}_{abcd}-\frac{1}{864}\bar{\theta}
                   (\Gamma_{ab}^{\sp \sp cdef}\hat{F}_{cdef}+24\Gamma^{cd}
                   \hat{F}_{abcd})\epsilon(\Gamma^{ab}
                   \theta)^{\mu} \nonumber \\
               & & +{\cal O}(\theta ^3), \\ 
 \Lambda _b^{
  \sp a}       &=& \lambda ^a_{\sp b}-\bar{\theta }\Gamma ^n\epsilon 
                   \hat{\omega }_{n \sp b}^{\sp a}+\frac{1}{144}\bar{\theta }
                   (\Gamma ^{a\sp rstu}_{\sp b}\hat{F}_{rstu}+24\Gamma _{rs}
                   \hat{F}_{\sp b}^{a \sp rs})\epsilon
                   +{\cal O}(\theta ^2), \\
 \Xi _{mn}     &=& \xi _{mn}-(\bar{\theta }\Gamma ^p\epsilon C_{pmn}+
                   \bar{\theta }\Gamma _{mn}\epsilon ) 
                   +\bar{\theta }\Gamma ^k\epsilon \bar{\theta }\Gamma^l
                   \psi _kC_{lmn}+\bar{\theta }\Gamma ^k\epsilon \bar{
                   \theta }\Gamma _{mn}\psi _k  \nonumber \\
               & & +\frac{4}{3}\bar{\theta}
                   \Gamma^l\epsilon\bar{\theta}
                   \Gamma_{l[m}\psi_{n]}+\frac{4}{3}\bar{\theta}\Gamma^l
                   \psi_{[n}\bar{\theta}
                   \Gamma_{|l|m]}\epsilon
                   +{\cal O}(\theta ^3), \\
 \Xi _{m\mu }  &=& \frac{1}{6}\bar{\theta}\Gamma^n\epsilon(\bar{\theta}
                   \Gamma_{mn})_{\mu}+\frac{1}{6}(\bar{\theta}\Gamma^n)_{\mu}
                   \bar{\theta}\Gamma_{mn}\epsilon+{\cal O}(\theta ^3), \\
 \Xi _{\mu 
        \nu }  &=& {\cal O}(\theta ^3) ,\\
 E_m^{\sp a}   &=& e_m^{\sp a}+2\bar{\theta }\Gamma ^a\psi _m -\frac{1}{4}
                   \bar{\theta }\Gamma^{acd}\theta \hat{\omega }_{mcd}+
                   \frac{1}{72}\bar{\theta }\Gamma _m^{\sp rst}\theta \hat{F}
                   _{rst}^{\sp \sp \sp a} \nonumber \\
               & & +\frac{1}{288}\bar{\theta }\Gamma ^{rstu}\theta \hat{F}
                   _{rstu}e_m^{\sp a}-\frac{1}{36}\bar{\theta }\Gamma 
                   ^{astu}\theta \hat{F}_{mstu}+{\cal O}(\theta ^3), \\
 E_m^
 {\sp \alpha}  &=& \psi _m^{\sp \alpha}-\frac{1}{4}\hat{\omega }_{mab}(\Gamma 
                   ^{ab}\theta )^{\alpha }+(T_m^{\sp rstu} \theta )^{\alpha}
                   \hat{F}_{rstu}   \nonumber \\
               & & +\bar{\theta}\Gamma^k\psi_m(T_k^{\sp abcd}
                   \theta)^{\alpha}\hat{F}_{abcd}-\frac{1}{576}\bar{\theta}
                   (\Gamma_{ab}^{\sp \sp cdef}\hat{F}_{cdef}+24\Gamma^{cd}
                   \hat{F}_{abcd})\psi_m(\Gamma^{ab}
                   \theta)^{\alpha}  \nonumber \\
               & & -12(T_k^{abcd}\theta)^{\alpha}\bar{\theta}\Gamma_{[ab}
                   \hat{D}_c\psi_{d]}-\frac{1}{4}(\bar{\theta}
                   \Gamma_a\hat{D}_m\psi_b-\bar{\theta}
                   \Gamma_b\hat{D}_m\psi_a+\bar{\theta}
                   \Gamma_m\hat{D}_a\psi_b)(\Gamma^{ab}
                   \theta)^{\alpha} \nonumber \\
               & &  +{\cal O}(\theta ^3), \\ 
 E_{\mu }^{
 \sp a}        &=&  -(\Gamma ^a \theta)_{\mu }+{\cal O}(\theta ^3),\\
 E_{\mu }^{
 \sp \alpha}   &=& \delta _{\mu }^{\sp \alpha }  \nonumber \\
               & & -\frac{1}{3}(\Gamma^k\theta)_{\mu}(T_k^{\sp abcd}
                   \theta)^{\alpha}\hat{F}_{abcd}+\frac{1}{1728}
                   ((\Gamma_{ab}^{\sp \sp cdef}\hat{F}_{cdef}+24\Gamma^{cd}
                   \hat{F}_{abcd})\theta)_{\mu}(\Gamma^{ab}
                   \theta)^{\alpha} \nonumber \\
               & & +{\cal O}(\theta ^3), \\ 
  \Omega _
 {\mu b}^{
 \sp \sp a}    &=& \frac{1}{144}\{(\Gamma ^{a\sp rstu}_{\sp b}\theta )_{\mu }
                   \hat{F}_{rstu}+24(\Gamma _{rs}\theta )_{\mu }\hat{F}_
                   {\sp b}^{a \sp rs} \} +{\cal O}(\theta ^2),  \\
 \Omega _{mab} &=& \hat{\omega }_{mab}+2\bar{\theta }\{ e^n_{\sp a}e^k_
                   {\sp b}(-\Gamma _kD_{[m}\psi _{n]}+\Gamma _nD_{[m}
                   \psi _{k]}+\Gamma _mD_{[n}\psi _{k]})\} \nonumber \\
               & & +\frac{1}{72}\bar{\theta}(\Gamma_{ab}^{\sp \sp rstu}
                   \hat{F}_{rstu}+24\Gamma_{rs}\hat{F}_{ab}^{\sp \sp rs})
                   \psi_m +{\cal O}(\theta ^2),  \\
 B_{mnl}       &=& C_{mnl}-6\bar{\theta }\Gamma _{[mn}\psi _{l]}+\frac{3}{4}
                   \hat{\omega }_{[l}^{\sp \sp cd}\bar{\theta }\Gamma _{
                   mn]cd}\theta -\frac{3}{2}\hat{\omega }_{[lmn]}\theta ^2
                   \nonumber \\
               & & -\frac{1}{96}\bar{\theta }\Gamma _{mnl}^{\sp \sp \sp rstu}
                   \theta \hat{F}_{rstu}-\frac{3}{8}\bar{\theta }\Gamma _{
                   [l}^{\sp \sp rs}\theta \hat{F}_{|rs|mn]}-12\bar{\theta }
                   \Gamma _a\psi _{[m}\bar{\theta }\Gamma ^a_{\sp n}\psi _{l]}
                   \nonumber \\
               & & +{\cal O}(\theta ^3), \\
 B_{mn\mu }    &=& (\bar{\theta }\Gamma _{mn})_{\mu }+\frac{8}{3}\bar{\theta}
                   \Gamma^k\psi_{[m}(\bar{\theta}\Gamma_{|k|n]})_{\mu}+
                   \frac{4}{3}(\bar{\theta}\Gamma^k)_{\mu}\bar{\theta}
                   \Gamma_{k[m}\psi_{n]}+{\cal O}(\theta ^3), \\
 B_{m\mu \nu } &=& (\bar{\theta }\Gamma _{mn})_{(\mu }(\bar{\theta }\Gamma ^n
                   )_{\nu )}+{\cal O}(\theta ^3), \\
 B_{\mu \nu 
 \rho }        &=& (\bar{\theta }\Gamma _{mn})_{(\mu }(\bar{\theta }\Gamma ^m
                   )_{\nu }(\bar{\theta }\Gamma ^n)_{\rho )}
                   +{\cal O}(\theta ^3). \\
\end{eqnarray}
Because the flat geometry had been known, we include the $\theta ^3$ term 
in $B_{\mu \nu \rho }$ for completeness.

Up to first order in anticommuting coordinates, the superfield 
components was investigated by E. Cremmer and S. Ferrara~\cite{kap}.
$\Xi^{k(2)},E_M^{\sp k(2)},\Xi_{MN}^{(2)},B_{LMN}^{(2)}$ was 
investigated by B. de Wit, K. Peeters and J. Plefka~\cite{wit98}.
$E_M^{\sp \alpha (2)},\Xi^{\mu (2)}$ was investigated in ref. ~\cite{shibu}.

\section{Computation}
$\Lambda_{ab}^{\sp \sp (2)}$ is subject to the following equations,
\begin{eqnarray}
\epsilon_1^{\nu}
\partial_{\mu}
\partial_{\nu}
\Lambda_{2ab}
^{\sp \sp (2)}
-(1 
\leftrightarrow 2) &=& \bar{\epsilon_2}\Gamma^n\epsilon_1(\Gamma^k
                       \psi_n)_{\mu}\hat{\omega}_{kba} -(\Gamma^n
                       \epsilon_1)_{\mu}\bar{\psi_n}\Gamma^k\epsilon_2
                       \hat{\omega}_{kba} \nonumber \\
                   & & -\frac{1}{144}\bar{
                       \epsilon_2}\Gamma^n\epsilon_1\{(\Gamma_{ba}^{\sp 
                       \sp cdef}\hat{F}_{cdef}+24\Gamma^{cd}\hat{F}_{bacd}
                       )\psi_n\}_{\mu} \nonumber \\
                   & & +\frac{1}{144}(\Gamma^n\epsilon_1)_{\mu}
                       \bar{\psi_n}(\Gamma_{ba}^{\sp \sp cdef}\hat{F}_{cdef}
                       +24\Gamma^{cd}\hat{F}_{bacd})\epsilon_2 \nonumber \\
                   & & +\frac{1}{6}(\Gamma_{ba}^{\sp \sp
                       cdef}\epsilon_2)_{\mu}\bar{\epsilon_1}\Gamma_{cd}
                       \hat{D}_e\psi_f+4(\Gamma^{cd}\epsilon_2)_{\mu}\bar{\epsilon_1}
                       \Gamma_{[ba}\hat{D}_c\psi_{d]} \nonumber \\
                   & & +(\Gamma^k\epsilon_2
                       )_{\mu}2\bar{\epsilon_1}(-\Gamma_a\hat{D}_{[k}\psi_{b]}
                       +\Gamma_k\hat{D}_{[b}\psi_{a]}+\Gamma_b\hat{D}_{[k}
                       \psi_{a]}) \nonumber \\
                   & & -(1 \leftrightarrow 2) . 
\end{eqnarray}
However, if simply we drive the equation,
\begin{eqnarray}
\epsilon_1^{\nu}
\partial_{\mu}
\partial_{\nu}
\Lambda_{2ab}
^{\sp \sp (2)}     &=& (\bar{\epsilon_2}\Gamma^n)_{\nu}\epsilon_1^{\nu}
                       (\Gamma^k\psi_n)_{\mu}\hat{\omega}_{kba} -(\Gamma^n)_{
                       \mu \nu}\epsilon_1^{\nu}\bar{\psi_n}\Gamma^k\epsilon_2
                       \hat{\omega}_{kba} \nonumber \\
                   & & -\frac{1}{144}(\bar{
                       \epsilon_2}\Gamma^n)_{\nu}\epsilon_1^{\nu}\{(
                       \Gamma_{ba}^{\sp 
                       \sp cdef}\hat{F}_{cdef}+24\Gamma^{cd}\hat{F}_{bacd}
                       )\psi_n\}_{\mu} \nonumber \\
                   & & +\frac{1}{144}(\Gamma^n)_{\mu \nu}\epsilon_1^{\nu}
                       \bar{\psi_n}(\Gamma_{ba}^{\sp \sp cdef}\hat{F}_{cdef}
                       +24\Gamma^{cd}\hat{F}_{bacd})\epsilon_2 \nonumber \\
                   & & +\frac{1}{6}(\Gamma_{ba}^{\sp \sp
                       cdef}\epsilon_2)_{\mu}\epsilon_1^{\nu}(\Gamma_{cd}
                       \hat{D}_e\psi_f)_{\nu}+4(\Gamma^{cd}\epsilon_2)_{\mu}
                       \epsilon_1^{\nu}(\Gamma_{[ba}\hat{D}_c
                       \psi_{d]})_{\nu} \nonumber \\
                   & & +(\Gamma^k\epsilon_2
                       )_{\mu}2\epsilon_1^{\nu}(-\Gamma_a\hat{D}_{[k}\psi_{b]}
                       +\Gamma_k\hat{D}_{[b}\psi_{a]}+\Gamma_b\hat{D}_{[k}
                       \psi_{a]})_{\nu} , 
\end{eqnarray}
this equation is inconsistent because $\mu$ and $\nu$ in the left-hand 
side of it are antisymmetric but these in the right-hand side of it are not
antisymmetric. Thus we must add terms which are symmetric under 
interchanging the indices 1 and 2 in the right-hand side of this
equation. 
\begin{eqnarray}
\epsilon_1^{\nu}
\partial_{\mu}
\partial_{\nu}
\Lambda_{2ab}
^{\sp \sp (2)}     &=& \bar{\epsilon_2}\Gamma^n\epsilon_1(\Gamma^k
                       \psi_n)_{\mu}\hat{\omega}_{kba} -(\Gamma^n
                       \epsilon_1)_{\mu}\bar{\psi_n}\Gamma^k\epsilon_2
                       \hat{\omega}_{kba} \nonumber \\
                   & & -\frac{1}{144}\bar{
                       \epsilon_2}\Gamma^n\epsilon_1\{(\Gamma_{ba}^{\sp 
                       \sp cdef}\hat{F}_{cdef}+24\Gamma^{cd}\hat{F}_{bacd}
                       )\psi_n\}_{\mu} \nonumber \\
                   & & +\frac{1}{144}(\Gamma^n\epsilon_1)_{\mu}
                       \bar{\psi_n}(\Gamma_{ba}^{\sp \sp cdef}\hat{F}_{cdef}
                       +24\Gamma^{cd}\hat{F}_{bacd})\epsilon_2 \nonumber \\
                   & & +\frac{1}{6}(\Gamma_{ba}^{\sp \sp
                       cdef}\epsilon_2)_{\mu}\bar{\epsilon_1}\Gamma_{cd}
                       \hat{D}_e\psi_f+4(\Gamma^{cd}\epsilon_2)_{\mu}\bar{\epsilon_1}
                       \Gamma_{[ba}\hat{D}_c\psi_{d]} \nonumber \\
                   & & +(\Gamma^k\epsilon_2
                       )_{\mu}2\bar{\epsilon_1}(-\Gamma_a\hat{D}_{[k}\psi_{b]}
                       +\Gamma_k\hat{D}_{[b}\psi_{a]}+\Gamma_b\hat{D}_{[k}
                       \psi_{a]}) +\bar{\epsilon_2}\epsilon_1\frac{438}{7}
                       (\hat{D}_{[b}\psi_{a]})_{\mu} \nonumber \\
                   & & +\bar{\epsilon_2}\Gamma^{ijk}\epsilon_1\{\frac{1}{28}
                       (\Gamma_{bkj}\hat{D}_{[a}\psi_{i]}-(a \leftrightarrow 
                       b))_{\mu}-\frac{31}{42}(\Gamma_{kji}\hat{D}_{[b}
                       \psi_{a]})_{\mu}\} \nonumber \\
                   & & +\bar{\epsilon_2}\Gamma^{ijkl}\epsilon_1\{-\frac{1}{6}
                       (\Gamma_{balk}\hat{D}_j\psi_i)_{\mu}+\frac{1}{21}
                       (\Gamma_{blkj}\hat{D}_{[a}\psi_{i]}-(a \leftrightarrow 
                       b))_{\mu} \nonumber \\
                   & & -\frac{1}{6}(\Gamma_{lkji}\hat{D}_{[b}
                       \psi_{a]})_{\mu}-\frac{17}{84}(\delta_{la}\Gamma_{kj}
                       \hat{D}_{[b}\psi_{i]}-(a \leftrightarrow b))_{
                       \mu} \nonumber \\
                   & & +\frac{1}{3}(\delta_{la}
                       \delta_{kb}-(a \leftrightarrow b))(\hat{D}_j\psi_i
                       )_{\mu}\} . 
\end{eqnarray}

Thus we obtain
\begin{eqnarray}
\Lambda_{ab}
^{\sp \sp (2)}     &=& \bar{\theta}\Gamma^n\epsilon \bar{\theta}\Gamma^k
                       \psi_n \hat{\omega}_{kba}-\frac{1}{144}\bar{\theta}
                       \Gamma^n\epsilon \bar{\theta}(\Gamma_{ba}^{\sp \sp cdef}
                       \hat{F}_{cdef}+24\Gamma^{cd}\hat{F}_{bacd})
                       \psi_n   \nonumber \\
                   & &  + \frac{1}{64}[\bar{\theta}\theta(-\frac{14016}{7}
                        \bar{\epsilon}\hat{D}_{[b}\psi_{a]}) \nonumber \\
                   & & +\frac{1}{6}\bar{\theta}\Gamma^{xyz}\theta\{-
                       \frac{48}{7}\bar{\epsilon}(\Gamma_{bzy}\hat{D}_{[a}
                       \psi_{x]}-(a \leftrightarrow b))+\frac{992}{7}\bar{
                       \epsilon}(\Gamma_{zyx}\hat{D}_{[b}\psi_{a]})\} \nonumber \\
                   & & +\frac{1}{24}\bar{\theta}\Gamma^{wxyz}\theta\{128
                       \bar{\epsilon}\Gamma_{bazy}\hat{D}_x\psi_w-
                       \frac{256}{7}\bar{\epsilon}(\Gamma_{bzyx}\hat{D}_{[a}
                       \psi_{w]}-(a \leftrightarrow b)) \nonumber \\
                   & & +128\bar{\epsilon}\Gamma_{zyxw}\hat{D}_{[b}\psi_{a]}
                       +\frac{1088}{7}\bar{\epsilon}(\delta_{za}\Gamma_{yx}
                       \hat{D}_{[b}\psi_{w]}
                       -(a \leftrightarrow b))\nonumber \\
                   & & -256\bar{\epsilon}(\delta_{za}\delta_{yb}\hat{D}_x
                       \psi_w-(a \leftrightarrow b))\}] .
\end{eqnarray}
$\Omega_{\mu ab}^{\sp \sp (2)}$ is subject to the following equation,
\begin{eqnarray}
\epsilon^{\nu}
\partial_{\nu}
\Omega_{\mu 
ab}^{\sp \sp (2)} &=& \frac{1}{144}\bar{\theta}\Gamma^n \epsilon \bar{\psi_n}
                      (\Gamma_{ba}^{\sp \sp cdef}\hat{F}_{cdef}+24\Gamma^{cd}
                      \hat{F}_{bacd})_{\mu}-\bar{\theta}\Gamma^n \epsilon 
                      (\Gamma^k\psi_n)_{\mu}\hat{\omega}_{kba} \nonumber \\
                  & & -(\Gamma^n
                      \epsilon)_{\mu}\bar{\theta}\Gamma^k\psi_n\hat{
                      \omega}_{kba}-(\Gamma^n\epsilon)_{\mu}2\bar{
                      \theta}(\Gamma_b\hat{D}_{[a}\psi_{n]}-\Gamma_a
                      \hat{D}_{[b}\psi_{n]}-\Gamma_n\hat{D}_{[b}
                      \psi_{a]}) \nonumber \\
                  & & -\frac{1}{144}(\Gamma^n\epsilon)_{\mu}\bar{
                      \theta}(\Gamma_{ab}^{\sp \sp cdef}\hat{F}_{cdef}
                      +24\Gamma^{cd}\hat{F}_{abcd})\psi_n +
                      \partial_{\mu}\Lambda_{ab}^{\sp (2)} .
\end{eqnarray}
Thus we obtain 
\begin{eqnarray}
\Omega_{\mu 
ab}^{\sp \sp (2)}  &=& -\frac{1}{64}[\bar{\theta}\theta\frac{14212}{7}
                       (\hat{D}_{[b}\psi_{a]})_{\mu} \nonumber \\
                   & & +\frac{1}{6}\bar{\theta}\Gamma^{xyz}\theta\{-
                       \frac{120}{7}(\Gamma_{bzy}\hat{D}_{[a}\psi_{x]}
                       -(a \leftrightarrow b))_{\mu}-\frac{1020}{7}(
                       \Gamma_{zyx}\hat{D}_{[b}\psi_{a]})_{\mu} \nonumber \\
                   & & -192(\delta_{za}\Gamma_b\hat{D}_y\psi_x
                       -(a \leftrightarrow b))_{\mu}-48(\delta_{za}\Gamma_y
                       \hat{D}_{[b}\psi_{x]}-(a \leftrightarrow b))_{
                       \mu}\}     \nonumber \\
                   & & +\frac{1}{24}\bar{\theta}\Gamma^{wxyz}\theta\{
                       \frac{32}{7}(\Gamma_{bzyx}\hat{D}_{[a}\psi_{w]}
                       -(a \leftrightarrow b))_{\mu}-132(\Gamma_{zyxw}
                       \hat{D}_{[b}\psi_{a]})_{\mu} \nonumber \\
                   & & -256(\delta_{za}\Gamma_{by}\hat{D}_x\psi_w
                       -(a \leftrightarrow b))_{\mu}+\frac{32}{7}(
                       \delta_{za}\Gamma_{yx}\hat{D}_{[a}\psi_{w]}
                       -(a \leftrightarrow b))_{\mu}\}] .
\end{eqnarray}

\section{Discussion}
We have obtained $\Lambda_{ab}^{\sp (2)},\Omega_{\mu a}^{\sp \sp b
(2)}$. Up to second order in anticommuting coordinates, only 
$\Omega_{m a}^{\sp \sp b(2)}$ remains. This component is complicated.  
However, it contains $\delta_{susy}\hat{D}_{[m}\psi_{n]}$ thus it is
expected to contain curvature terms. From Bianchi identity, curvature
terms should appeared in vielbein superfields at third and the higher order in 
anticommuting coordinates. This gives interaction terms coupled to
curvature in Matrix model and higher curvature corrections to Einstein 
gravity in low energy effective actions. Thus it is important that 
we investigate $\Omega_{m a}^{\sp \sp b(2)}$ . These terms and 
terms which are required to
obtain terms of Matrix theory which are third order in anticommuting 
coordinates is under considerations.  

However, $\Omega_{\mu a}^{\sp \sp b(2)}$ is also important in study of 
superspace structure and curved membrane action. $\kappa$-symmetry 
constraints act on torsion fields and curvature fields. Torsion are 
defined as $T^A = DE^A =dE^A + E^B \Omega_B^{\sp A}$. Thus 
$\Omega_{\mu a}^{\sp \sp b(2)}$ has information about more higher
components of vielbein than $\Omega_{m a}^{\sp \sp b(2)}$. 
Moreover, $\Omega_{\mu a}^{\sp \sp b(2)}$ contains
$\hat{D}_{[a}\psi_{b]}$ which is nonlinearly exact supersymmetry field 
strength which information is important in study of supersymmetry.

\section*{Acknowledgments}

I would like to thank Y.Matsuo for valuable suggestions. 
\appendix
\section*{Appendix}
\section{Conventions}
\subsection{Indices}

We use Greek indices for spinorial components and Latin indices for
vector components. And we use former alphabet for the tangent space
indices and later for general coordinates indices: $a,b,c,...$ for tangent
vector indices and $k,l,m,...$ for general vector indices,
and  $\alpha ,\beta ,...$ for tangent spinorial indices and 
$\mu , \nu ,...$ for general spinorial indices.

Superspace coordinates $(x^m ,\theta ^{\mu })$ are designated 
$Z^M $ , where later capital Latin alphabet $M,N,..$ are collective 
designations for general coordinate indices. While former capital
Latin alphabet $A,B,..$ are collective designations for tangent
space indices.

\subsection{p-form superfield}

We introduce p-form superfields as follows,
\begin{eqnarray}
  X               &\equiv & \frac{1}{p!} dz^{M_p}...dz^{M_1} 
                            X_{M_p ... M_1} \nonumber \\
                  &\equiv & \frac{1}{p!} E^{A_p}...E^{A_1} X_{A_p ... A_1}, \\
  X_{A_p ... A_1} &\equiv & \sum_{i=1}^{32} X_{A_p ... A_1}^{\sp \sp \sp (i)}.
\end{eqnarray}
 $X_{A_p ... A_1}^{\sp \sp \sp (i)}$ is component at i-th order in 
anticommuting coordinates.

\subsection{Brackets}

Symmetrization bracket $(\sp \sp )$ and antisymmetrization bracket
$[\sp \sp ]$ is defined as follows,

\begin{eqnarray}
   [M_1 ... M_N]   &=& \frac{1}{N!}( M_1 ... M_N \sp +
                               \mbox{antisymmetric terms} ), \nonumber \\
   (M_1 ... M_N)   &=& \frac{1}{N!}( M_1 ... M_N \sp +
                               \mbox{symmetric terms} ).
\end{eqnarray}
 
\subsection{Gamma matrices(11-dimensional)}

Since we use the Majorana representation, all components are real.

Gamma matrix $\Gamma^{a \sp \alpha}_{\sp \sp \sp \beta}$ is
defined as follows,
\begin{eqnarray}
 \{ \Gamma^{a} ,\Gamma^{b} \} = 2 \eta ^{ab}.
\end{eqnarray}
We use the mostly plus metric; $\eta_{ab}\sim (-+...+)$.
We lower the spinorial indices
by charge conjugation matrix $C_{\alpha \beta }$.
\begin{eqnarray}
  \bar{\psi}_{\beta} = \psi ^{\alpha }C_{\alpha \beta }, \nonumber \\
  \Gamma^a _{\sp \sp \alpha \beta}=C_{\alpha \gamma }
  \Gamma^{a \sp \gamma}_{\sp \sp \sp \beta} .
\end{eqnarray}
$\Gamma^{a_1..a_n}_{\sp \sp \sp \sp \alpha\beta}(n=1,2,5,6,9,10)$ are 
symmetric matrices and $\Gamma^{a_1..a_n}_{\sp \sp \sp \sp \alpha\beta}
(n=0,3,4,7,8,11)$ are antisymmetric matrices.

%\bibliographystyle{prsty}
%\bibliography{phys}

\end{document}